\documentclass[manuscript]{aastex}
\usepackage{graphicx}
\usepackage{color}
\begin{document}

\shorttitle{Multi-wavelength Emission .....}
 \shortauthors{Cheng et al.}

\title{Multi-wavelength Emission from the Fermi Bubble II. Secondary Electrons and the Hadronic Model of the Bubble.}
\author{K.-S. Cheng$^{1}$, D. O. Chernyshov$^{1,2}$, V. A. Dogiel$^{1,2,4}$, and  C.-M.
Ko$^{3}$}
 \affil{$^1$Department of Physics, University of Hong
Kong, Pokfulam Road, Hong Kong, China}
\affil{$^2$I.E.Tamm Theoretical Physics Division of P.N.Lebedev Institute of Physics, Leninskii pr. 53, 119991 Moscow, Russia}
\affil{$^3$Institute of
Astronomy, Department of Physics and Center for Complex Systems,
National Central University, Jhongli, Taiwan}
\affil{$^4$Moscow
Institute of Physics and Technology (State University), 9,
Institutsky lane, Dolgoprudny, 141707, Russia}



\altaffiltext{0}{........}

\begin{abstract}
We analyse the origin of the gamma-ray flux from the Fermi Bubbles (FBs)
in the framework of the hadronic model in which  gamma-rays are
produced by collisions of relativistic protons with the protons of background plasma in
the Galactic halo. It is  assumed in this model that the observed radio
emission from the FBs is due to synchrotron radiation of secondary
electrons  produced by $pp$ collisions. However, if
these electrons loose their energy by the synchrotron and inverse-Compton,
the spectrum of secondary electrons is too soft, and an additional
arbitrary component of primary electrons is necessary in order to reproduce
the radio data. Thus, a mixture of the hadronic and leptonic models is required for the observed radio flux. It was shown that if the spectrum
of primary electrons is $\propto E_e^{-2}$, the permitted range of
the magnetic field strength is within 2 - 7 $\mu$G region. The
fraction of gamma-rays produced by $pp$ collisions can reach about
80\% of the total gamma-ray flux from the FBs. If magnetic field is $<2$ $\mu$G or $>7$ $\mu$G the model is
unable to reproduce the data. Alternatively, the electrons in the
FBs may lose their energy by adiabatic energy losses if there is
a strong plasma outflow in the GC. Then, the pure hadronic
model is able to reproduce characteristics of the radio and
gamma-ray flux from the FBs. However, in this case  the required
  magnetic field strength in the FBs and the power of CR
sources are much higher than those followed from observations.
\end{abstract}


\date{\today}

\maketitle

\section{Introduction}
The origin of  giant gamma-ray structure  named as Fermi
bubbles (FBs) discovered from  analysis of the Fermi-LAT data
\citep[see,][]{dob10,meng,acker14}  is still enigmatic. Two main
interpretations of gamma-ray production in the bubbles were
suggested which can be defined as hadronic and leptonic. In the
latter case gamma-rays are produced by inverse Compton scattering of
relativistic electrons on background photons \citep[see
e.g.][]{meng}. Because of relatively short lifetime of
relativistic electrons they should be produced inside or nearby
regions of emission. The in-situ acceleration can be provided
either by a shock or shocks which are generated by tidal processes
nearby the central black hole  or by an MHD-turbulence which is
excited behind a shock \citep[for some aspects of the leptonic
model see e.g.][and others]{cheng,cheng14a,mertsch,yang12}.

Alternatively, the hadronic model of the FBs was suggested and
developed in a series of publications by
\citet{crock11,crocker11,crock13,fuji,thoudam13,Yang2014}. This
model does not require in-situ acceleration of protons because of
their relatively long lifetime.  Thus, \citet{crock11} concluded
from observational data that star formation regions  within the
central 200 pc radius  release energy continuously, with the
average power $\sim 10^{40}$ erg s$^{-1}$. They assumed that this
energy is transformed into a flux of  relativistic cosmic rays
(CRs) which are carried away into the halo by a plasma outflow
from the Galactic Center (GC). They assumed also that CR protons
fill the bubble region and are trapped somehow there  for the time
of $pp$ collisions, $\tau_{pp}\sim (n_H\sigma_{pp}c)^{-1}\sim
10^{10}$ yr, where $\sigma_{pp}$ is the cross-section of $pp$
collisions and $n_H\simeq 10^{-2}$cm$^{-3}$ is the plasma density
in the halo.

According to \citet{Yang2014} the leptonic model  is unable
to  interpret the observed hardening of the gamma ray spectrum to
the FB edges that is a problem for the leptonic model. We suppose,
however, that special investigations of this effect are necessary
in order to get a more reliable conclusion.

The hadronic scenario requires  a rather small diffusion
coefficient of protons  within the bubble, $\sim 10^{26}$
cm$^2$s$^{-1}$ at 1 GeV \citep[see][]{crock11}, which is two
orders of magnitude smaller than in the Galaxy, or some sort of "magnetic walls" near the FB edges is needed to confine protons in the FBs for a long time \citep[see][]{jones12}.

 In these assumptions a power in CR
proton about $W\simeq 2\times 10^{38}$ erg s$^{-1}$ is required in order to produce the gamma-ray
flux $F_\gamma \simeq 4\times 10^{37}$ erg s$^{-1}$ from the FBs. Such a power can easily be supplied
by star formation regions in the GC. The spectrum of gamma-rays from the FBs can be presented as power-law,
\begin{equation}
F_{\gamma}^{FB}\propto E_\gamma^{-2}\,,
\label{gfb}
\end{equation}
  in the range $E_\gamma =1 - 100$ GeV \citep[see][]{meng}.
\citet{crock11} estimated a radio flux from FBs produced by
secondary electrons generated by $pp$ collisions. They obtained
that this radio flux in the range 20-60 GHz is about
$\Phi_\nu\simeq 2\times 10^{36}$ erg s$^{-1}$ for the magnetic
field strength is $H>10$ $\mu$G that is about of the flux
derived by \citet{fink} from the WMAP data, $\Phi_\nu\simeq
(1-5)\times 10^{36}$ erg s$^{-1}$. These and subsequent
observations of radio emission from the FBs
\citep[see][]{jones12,ade13} showed  that the FB radio spectrum is
power-law:
\begin{equation} \Phi_\nu^{FB}\propto \nu^{-\alpha}\,,
\label{rfb}
\end{equation}
where $\alpha$ is the spectral index of radio emission which ranges from 0.5 to 0.63 at GHz frequencies. If the emission is synchrotron emission, then spectrum of the radio emitting electrons is close to a power-law,
\begin{equation}
N_e \propto E_e^{-(2\alpha+1)}\,.
\label{nefb}
\end{equation}

Below we would like to derive the parameters of the hadronic model  at which the main part of
radio and gamma-ray emission from the FB, can be produced, indeed, by proton interaction with
the gas in the Galactic Halo (hadronic model).

\section{Spectra of gamma-rays and secondary electrons in the hadronic model}

In the hadronic model, when protons are trapped in  the FBs
and lose   energy in  $pp$ collisions, their spectrum, $N_p(E)$,
can be described  in the framework of the "leaky box model"
\citep[see for details]{ber90}
\begin{equation}
n_H\sigma_{pp}cN_p(E) = Q(E)
\label{source}
\end{equation}
where  the constant $Q(E)$ is  the power of proton sources in the GC.

 The intensity of $pp$ gamma-ray emission can be calculated from
\begin{equation}
I_{pp}(E_\gamma) =
\frac{n_HcL}{4\pi}\int \frac{d\sigma_{pp}(E,E_\gamma)}{dE_\gamma}N_p(E)dE ,
\label{ipp}
\end{equation}
where $d\sigma_{pp}/dE_\gamma$ is the differential cross-section
of production of gamma-rays  \citep[see e.g.][]{kamae,shiba13},
and $L$ is the thickness of radiating region. Below we calculate intensities of gamma-ray and microwave emission at relatively high latitudes where the densities
of the gas, photons,  magnetic field strength and CR protons are supposed to vary
relatively slowly along the path of view. Therefore, we neglect their
spatial variations in the halo and use average values of 
these components there.

If the FB gamma-ray flux is generated by  $pp$ collisions of
relativistic protons with protons of background plasma, then as
follows from Eq. (\ref{gfb}) the spectrum of protons needed to
reproduce the flux of gamma-rays from the FBs is
\citep[see][]{crock11}
\begin{equation}
N_p(E)=K_p  E^{-2}\theta\left(E^p_{max}-E\right)\,,
\label{pfb}
\end{equation}
where $K_p$ is a constant and $E^p_{max}$ is the maximum energy of emitting protons whose value can be estimated from the cut-off position, $E_\gamma^{cut-off}$, in the FB gamma-ray spectrum. According to the recent data analysis of \citet{acker14} the cut-off position is $E_\gamma^{cut-off}\leq 200$ GeV that gives the estimate for  $E^p_{max} \simeq 3$ TeV \citep[see][]{atoyan}.

The production spectrum of secondary electrons is described by the
equation
\begin{equation}
Q(E_e) = n_Hc\int \frac{d\sigma(E,E_e)}{dE_e}N_p(E)dE ,
\label{eqfb}
\end{equation}
where $d\sigma/dE_e$ is the differential cross-section of
secondary electron production  by $pp$ collisions \citep[see
e.g.][]{kamae,shiba13}.

In the framework of the "leaky box model" the spectrum of secondary electrons in the FBs is
\begin{equation}
N_{se}(E_e) =\frac{1}{|dE/dt|}\int\limits_{E_e}^\infty Q(E_e^\prime)dE_e^\prime
\label{se}
\end{equation}
where $dE/dt$ is the rate of electron energy losses.
Relativistic electrons in the halo lose their  energy by
synchrotron and inverse-Compton \citep[see][]{blum}, and for for electrons
with sufficiently low energies the losses can be approximated as
\begin{equation}
\frac{dE}{dt} = -c\sigma_T\left(w_{ph}+\frac{H^2}{8\pi}\right)\left(\frac{E}{mc^2}\right)^2=-\beta(H) E^2\,,
\label{lefb}
\end{equation}
where $\sigma_T$ is the Thomson cross-section,  $w_{ph}$ is
the total density of  background photons in the central part of
the Galactic halo which is taken about 2 eV cm$^{-3}$\citep[see
e.g.][]{acker11}, $H$ is the magnetic field strength whose value
will be derived below.

The energy of  secondary electron, $E_{e}$, is proportional to the energy of primary proton, $E_p$ as  $E_{e}\simeq 0.039E_p$. Then the production rate of secondary electron for energies of protons above the reaction theshold can be presented as \citep[][see for details]{atoyan}
\begin{equation}
Q(E_e) \simeq \frac{n_Hc\sigma_{pp}}{0.039}N_p(E_e/0.039)\,.
\end{equation}
From Eq. (\ref{pfb}) we obtain
\begin{equation}
N_{se}(E_e) \simeq E_{e}^{-3}\frac{K_pn_Hc\sigma_{pp}}{0.039\beta(H)} = K_{se}E_e^{-3}
\end{equation}

These electrons contribute also into the total flux of gamma-rays
from the FBs by inverse-Compton scattering. The gamma-ray intensity in the
direction of the Galactic coordinates $(\ell,b)$ contributed by
the FB is
\begin{equation}
I_\gamma(E_\gamma,\ell,b)=\frac{c}{4\pi}\int\limits_{L(\ell,b)}d
L\int\limits_\epsilon
n(\epsilon,r)d\epsilon\int\limits_{p}p^2f(r,p,t)\left(\frac{d^2\sigma}{d\epsilon~dp}\right)_{KN}dp\,.
\label{gamma}
\end{equation}
where $L(\ell,b)$ is the line of sight in the direction
$(\ell,b)$,  $\sigma_{KN}$ is the Klein-Nishina cross-section
\citep[see][]{blum}, $\epsilon$  is the energy of background
photons and $n(\epsilon)$ is their density.

 Below we define parameters of which the hadronic model can be
reanalyzed.

\section{Origin of the Radio and Gamma-Ray Emission from the FB in the Hadronic Model
with an Arbitrary Flux of Primary Electrons}

Because the characteristic time of electron energy losses is much
shorter than the lifetime  of CR in the bubble ($10^{10}$ yr),
from Eqs. (\ref{pfb}) - (\ref{lefb}) it follows that the spectrum
of secondary electrons in the FBs is
\begin{equation}
N_{se}\propto E_e^{-3}
\end{equation}
that is softer than required by the radio data, see Eq.
(\ref{nefb}). Pure hadronic model is unable  to reproduce the
FB radio spectrum if the lifetime of secondary electrons is
determined by the synchrotron and inverse Compton energy losses.
Therefore, an additional component of "primary" electrons with a
hard spectrum is necessary in order to reproduce the observed
radio emission from the FBs.

Because the spectral index of microwave spectrum is around
0.5 (see Eq.(\ref{rfb})), the  spectrum of primary electrons
cannot be steeper than $E_e^{-2}$. Therefore we take it  in the
arbitrary form
\begin{equation}
N_{pe}(E_e) = K_{pe}E_e^{-2}\theta(E_{max}- E_e).
\end{equation}
where $\theta(x)$ is the Heaviside function.

The intensity of synchrotron emission in the direction ${\bf
\ell}$ is described by the  equation \citep[see for
details]{ginz79}
\begin{equation}
I(\nu)=\frac{\sqrt{3}e^3HL}{mc^2}\int\limits_0^{E_{max}}N(E_e,)dE_e\frac{\nu}{\nu_c}\int
\limits_{\nu/\nu_c}^\infty K_{5/3}(x)dx
\end{equation}
where $N(E_e)$ is the density of electrons with the energy $E_e$,
$K_\alpha(x)$ is the McDonald function a nd $\nu_c=
3eH\gamma^2/4ðmc$, $\gamma$ is the Lorenz-factor of an electron.

Below for estimates we use approximations for synchrotron emission
of primary and secondary  electrons in the form \citep[][see for
details]{ginz79}
\begin{eqnarray}
&&I_{p}(\nu)= K_{pe} 0.103\frac{e^3}{mc^2}\left(\frac{3e}{4\pi m^3c^5}\right)^{0.5}H^{3/2}L\nu^{-0.5}=K_{pe} I_p^0(\nu)\,,\nonumber\\
&&I_{s}(\nu)= K_{se} 0.074\frac{e^3}{mc^2}\left(\frac{3e}{4\pi m^3c^5}\right)H^{2}L\nu^{-1}=K_{se} I_s^0(\nu)\,.
\label{synchr}
\end{eqnarray}

 For estimates we take the microwave  spectrum from \citet{ade13} which gives the intensity $I$ 
\begin{eqnarray}
&&4.3\cdot 10^{-19}\leq I_{23} \leq
4.6\cdot 10^{-19} ~\mbox{erg cm}^{-2}\mbox{s}^{-1}\mbox{sr}^{-1}\mbox{ at 23 GHz}\,,\label{radio_ineq01}\\
&&2.4\cdot 10^{-19}\leq I_{61}\leq 2.7\cdot 10^{-19}~\mbox{erg cm}^{-2}\mbox{s}^{-1}\mbox{sr}^{-1}\mbox{ at 61 GHz}\,.
\label{radio_ineq02}
\end{eqnarray}
Then we have
\begin{eqnarray}
&&I_{23}=K_{se} I_s^{0}(23 GHz) +K_{pe} I_{p}^{0}(23
GHz)\,,\\
&&I_{61} =K_{se} I_{s}^{0}(61 GHz) +K_{pe} I_{p}^{0}(61 GHz)\,,
\end{eqnarray}
which give
\begin{eqnarray}
&&K_{se}(H)=\frac{I_{23}I_{p}^{0}(61)-I_{61}I_{p}^{0}(23)}{I_{s}^{0}(23)I_{p}^{0}(61)-I_{s}^{0}(61)I_{p}^{0}(23)}\\
&&K_{pe}(H)=\frac{I_{23}I_{s}^{0}(61)-I_{61}I_{s}^{0}(23)}{I_{p}^{0}(23)I_{e}^{0}(61)-I_{p}^{0}(61)I_{e}^{0}(23)}
\end{eqnarray}

The magnetic field strength $H$ can be estimated from the measured
intensity of gamma rays from the  FBs \citep[see][]{meng} which is
$E_\gamma^2I_\gamma(E_\gamma)\simeq 4.2\times 10^{-7}$ GeV
cm$^{-2}$s$^{-1}$sr$^{-1}$.  From the approximations formulas for
proton gamma-ray production by $pp$ collisions
\citep[see][]{atoyan} and by inverse Compton scattering of primary
electrons on background photons \citep[see][]{ginz79} the equation
for the total gamma-ray emission is (here we neglected the
contribution from secondary electrons)
\begin{equation}
E_\gamma^2I_\gamma(E_\gamma)\simeq\frac{cL}{4\pi}\left(K_p(H)0.075n_H\sigma_{pp}+K_{pe}(H)\sigma_Tn_{ph}
\frac{(4/3\varepsilon_{ph}E_\gamma)^{0.5}}{mc^2}\right)
\label{gamma_FB}
\end{equation}

\begin{figure}[h]
\centering
\includegraphics[width=0.8\textwidth]{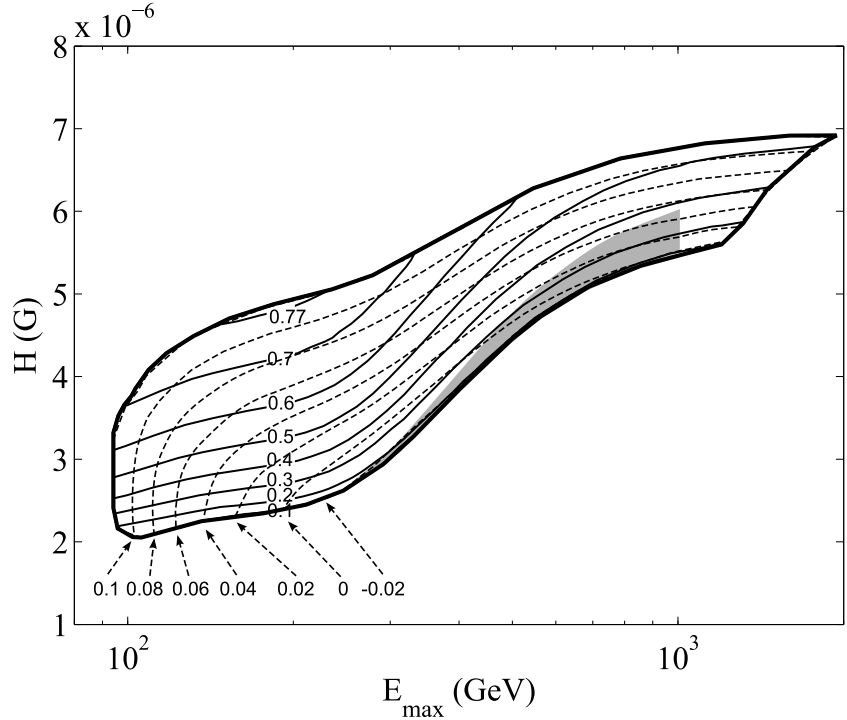}
\caption{Contours of $\chi_{pp}=const$ (solid lines) and  contours of deviations of the model spectral index $\alpha$ from $\bar{\alpha}=0.56$,   $\Delta\alpha=\alpha-\bar{\alpha}=const$ (dashed lines). The spectral index  $\bar{\alpha}=0.56$ is derived by \citet{ade13} from the Planck data.
The shaded area shows the region of parameters for the pure leptonic model.}  \label{contour}
\end{figure}

As an example we take the intensity of gamma-ray emission at $E_\gamma=1$ GeV which is produced by the inverse Compton 
  scattering on the relic photons
($\varepsilon_{ph}\simeq 6.6\times 10^{-4}$ eV, $n_{ph}\simeq 400$
cm$^{-3}$). For the most favourable parameters for the proton
contribution (the steepest spectrum of gamma-ray within error bars) we obtain from Eq.(\ref{gamma_FB})  that the required
magnetic field in the FBs in this case is $H\sim 1.2\times
10^{-5}$. The upper limit of contribution of protons into the
total flux is about 36\% while the rest 64\% are produced by
inverse Compton scattering of primary electrons on the relic
photons. These estimates were derived from Eq. (\ref{gamma_FB}) in which we used analytical approximations of gamma-ray production by electrons and protons with power-law spectra. Besides, the cross-sections of these processes were taken there as constants. More accurate results for arbitrary spectra of emitting particles can be obtained from numerical calculations with the cross-sections from \citet{blum} and \cite{kamae} which in this case are functions of particle energy. Below we define the fraction of the FB gamma-rays produced by protons as $\chi_{pp}$. From the  inequalities (\ref{radio_ineq01}) and (\ref{radio_ineq02}) we calculate numerically the maximum value of ${\bar\chi}_{pp}$ for given the magnetic field strength $H$ and the cutoff energy of primary electrons $E_{max}$ although other values of $\chi_{pp}<{\bar\chi}_{pp}$ are possible within the limits of the inequalities. The result of numerical calculation of a contribution from primary
protons is shown in Fig. \ref{contour}. The solid lines show the
maximum fraction of the FB gamma-rays produced by protons, ${\bar\chi}_{pp}$, as
a function of the magnetic field strength $H$ and the cutoff
energy of primary electrons $E_{max}$. 

We showed also in this figure variations of $\Delta\alpha=\alpha-\bar{\alpha}$ where $\alpha$ is the model spectral index of microwave emission derived for different $H$ and $E_{max}$ and $\bar{\alpha}=0.56$ is  obtained by \citet{ade13} from the Planck data. The range of model index $\alpha$  in the frequency interval from 23 to 61 GHz is restricted between 0.48 and 0.67 as follows  from inequalities (\ref{radio_ineq01})-(\ref{radio_ineq02}). From the figure we see that the higher is the value of $\alpha$, the larger is $\chi_{pp}$ (see Fig. \ref{contour}). 

The maximum of
protons contribution (for $\alpha=0.67$) is about 78\% if the
magnetic field strength is $H\simeq 5\times 10^{-6}$G  
that is about two times larger than follows from Eq. (\ref{gamma_FB}). The reasons for this discepancy are simplifications of analytical equations describing gamma-ray production as we mentioned above.

The thick solid line
shows the area of permitted parameters when the hadronic model reproduces
the FB microwave and gamma-ray fluxes. For each value of $E_{max}$ the range of permitted values of $H$ is restricted. Outside this area the model is unable to satisfy Eqs. (\ref{radio_ineq01})-(\ref{radio_ineq02})  simultaneously.  The shaded area shows the range of parameters when the origin of these fluxes is pure leptonic when the microwave spectrum is close to $\nu^{-0.5}$. The pure leptonic model corresponds to the condition $\chi_{pp}=0$. Within limits of the inequalities (\ref{radio_ineq01}) and (\ref{radio_ineq02}) the condition $\chi_{pp}=0$ is satisfied for different  $H$ and $E_{max}$ whose values are shown by the shaded area.

We notice that in our analysis we accepted higher deviations from $\bar{\alpha}$ than derived by \citet{ade13}, $\bar{\alpha}=0.56\pm 0.05$, that increases a permitted value of $\chi_{pp}$.

The magnetic field strength in the FBs is quite
uncertain. \citet{thoudam13}  estimated its value from the GALPROP
program that gave 1.3 $\mu$G  at the altitude 5 kpc above the
Galactic plane. The estimations of magnetic field inside the
Bubbles by \citet{jones12} and \citet{carr13} ranges from 6 to 15
$\mu$G.

\begin{figure}[h]
\centering
\includegraphics[width=0.8\textwidth]{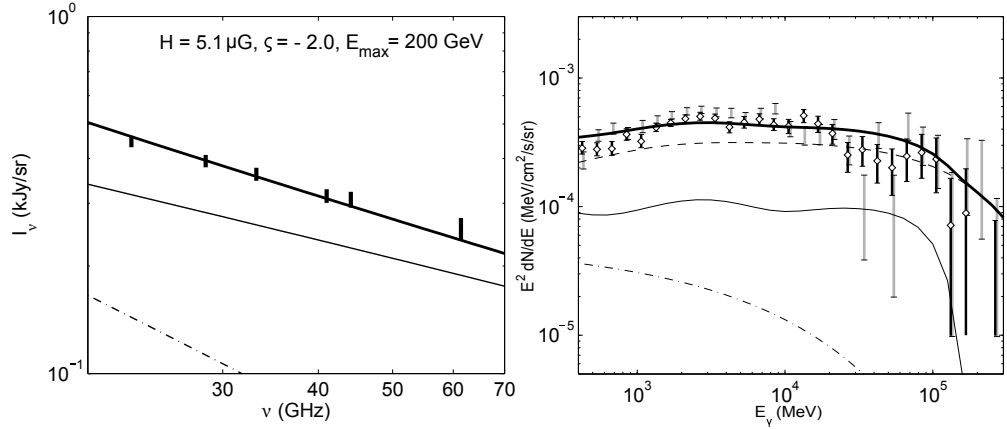}
\caption{Spectra of radio (left panel) and gamma-ray (right panel)
emissions from FB for $H = 5$ $\mu$G. Thin solid line -
contribution from primary electrons, dashed-dotted line -
contribution from secondary electrons and dashed line (right
panel) contribution of pp into the gamma-ray flux. The resulting
spectra are shown in both panels by the heavy solid lines. The
datapoints in the left panel were taken from \citet{ade13}  and in
the right panel   from \citet{hoop} for the latitude range
$20^\circ-30^\circ$ (gray lines) and for the range
$30^\circ-40^\circ$ (black lines with diamonds). }  \label{pp_max}
\end{figure}

The spectra of radio and gamma-ray emission for $H\sim 5$ $\mu$G
are shown in Fig. \ref{pp_max}. We see that even in the most
favourable case for the hadronic model (when the magnetic field
strength is about $5$ $\mu$G) the contribution from protons (pp
collisions + IC from secondary electrons/positrons) is about 80\%,
and the remaining 20\%  is produced by primary electrons. The
gamma-ray flux from $pp$ collisions is $3.2\times 10^{-9}$ erg
s$^{-1}$cm$^{-2}$sr$^{-1}$, from secondary electrons is $2.5\times
10^{-10}$ erg s$^{-1}$cm$^{-2}$sr$^{-1}$, and the necessary flux
from primary electrons is $1.1\times 10^{-9}$ erg
s$^{-1}$cm$^{-2}$sr$^{-1}$. Thus, we conclude that the pure
hadronic origin of the nonthermal emission (gamma and radio) from
the FBs is problematic.

\section{Effect of Adiabatic Energy Losses}

As we see from the previous section the pure hadronic model is
unable to  reproduce both radio and gamma-ray emission from the
FBs if secondary electrons lose their energy by synchrotron
and inverse  Compton. The radio spectrum of secondary electron is
too soft, and an additional arbitrary component of electrons with
a relatively hard spectrum.

However, as we showed in \citet{cheng} the situation is different
if adiabatic losses are significant in the FBs. For a divergent
outflow  from the GC region with the velocity ${\bf u}$ the rate
of adiabatic  energy losses is $dE/dt=-E\nabla\cdot {\bf u}/3$.

 There are arguments, indeed, in favour of plasma outflow from the Galactic
 central region \citep[see][]{crock11,carr13}. Numerical and analytical calculations of  \citet{breit91,breit}
showed that the outflow velocity in the halo increased linearly with the altitude $z$ above the Galactic Plane.
Thus, we can approximate the velocity distribution in the form $u(z)=3\lambda z$, and the rate of adiabatic
and synchro-Compton losses has the form $dE/dt=-(\lambda E+\beta E^2)$. From Eq. (\ref{se}) we obtain that
\begin{equation}
N_{se}=\frac{Q}{\lambda(\varsigma+1)}\frac{E^{-\varsigma}}{(1+\beta E/\lambda)}\,.
\label{ase}
\end{equation}
where the source function was taken in the form of equation (\ref{source}). We
see that for energies  $E<\lambda/\beta$ the spectrum of secondary
electrons is flat that is necessary for the observed radio
emission from the FBs and an additional component of primary
electrons is unnecessary in this case.

\begin{figure}[h]
\centering
\includegraphics[width=0.9\textwidth]{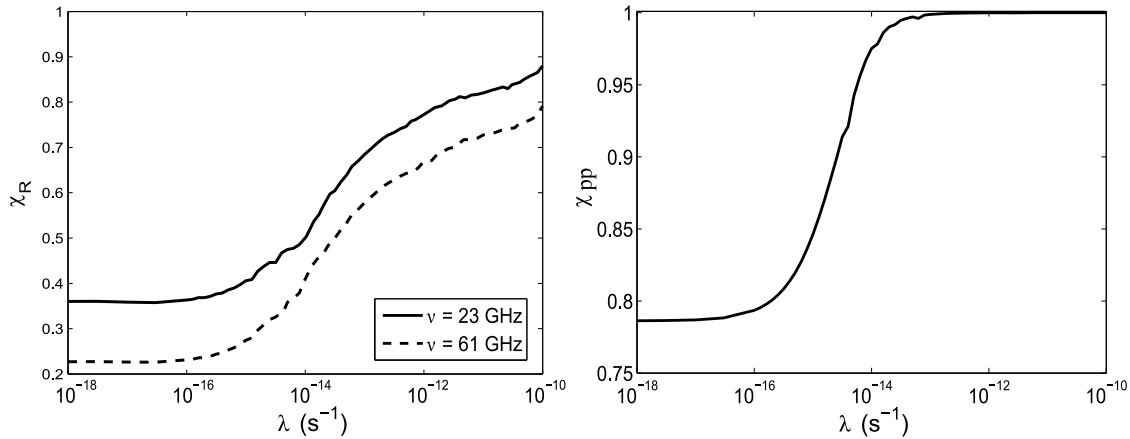}
\caption{{\it Left panel}. The fraction of radio flux from the FBs
at frequencies 23 and  61 GHz which is produced by the secondary
electrons as a function of the velocity gradient $\lambda$. The
total radio flux from the FBs was taken from \citet{ade13}. {\it
Right panel.} The fraction of gamma-ray flux from the FBs produced
by $pp$ collisions as a function of the velocity gradient
$\lambda$.}  \label{chiR}
\end{figure}

If the velocity gradient is relatively small, then a component of
primary electrons is still  required in order to reproduce the
radio data. However, our numerical calculations show that the
fraction of radio emission from secondary electrons increases with
the increase of the outflow gradient velocity $\lambda$ as shown in
Fig. \ref{chiR} (left panel). The value of $\chi_R$ in this
figure shows the fraction of the radio flux from the FBs produced
by the secondary electrons. The other part $(1-\chi_R)$ of the
flux is generated by primary electrons.

The fraction of the total gamma-ray flux produced by $pp$
collisions in the model with the  outflow is shown in Fig.
\ref{chiR} (right panel).
 As we see the hadronic model with adiabatic losses describes quite well the gamma and
radio emission from the FB for very large $\lambda$, and in the limit $\lambda \gg 10^{-10}$ s$^{-1}$  no
additional component of  primary electrons is needed.

\begin{figure}[h]
\centering
\includegraphics[width=0.9\textwidth]{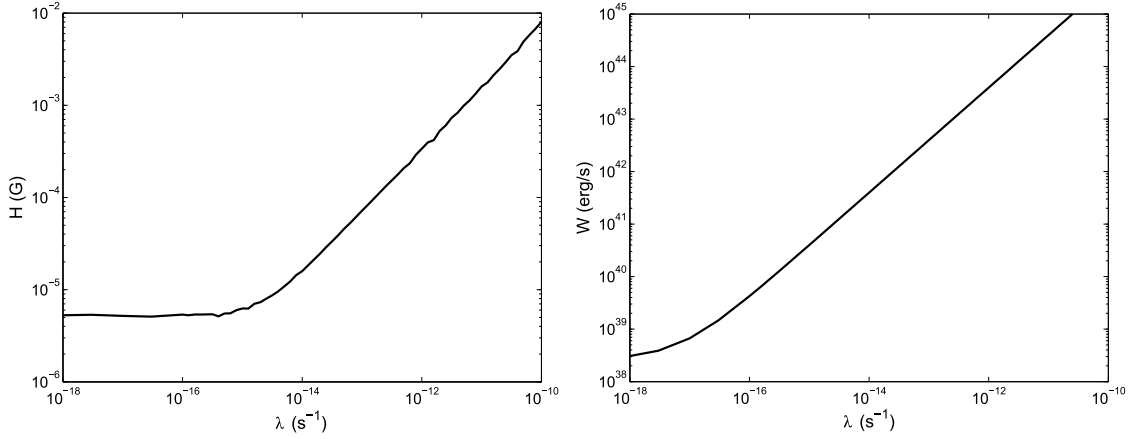}
\caption{{\it Left panel.} The strength of magnetic field
required for the observed radio flux from the FBs as a function of
the velocity gradient $\lambda$. {\it Right panel.} The power of CR
sources required for production of the gamma-ray flux from the FBs
by $pp$ collisions as a function of the velocity gradient
$\lambda$.}  \label{alphaH}
\end{figure}

However, the density of CRs drops down for high velocity gradients
$\lambda$ (see Eq. (\ref{ase})).  Therefore, for high $\lambda$ a
stronger magnetic field is necessary in order to produce the
observed radio flux from the FBs and a higher power of CRs sources
for the $pp$ gamma-ray flux from there. This effect is illustrated
in Fig. \ref{alphaH}  (left and right panels).

From these figures we see that the hadronic model with an outflow
requires unrealistically high  magnetic field strength and the CR
power in the FBs in order to reproduce the observed gamma and
radio fluxes from there. In our opinion this is a serious problem
of the model.

\section{Conclusion}
The conclusions can be itemized as follows:
\begin{itemize}
\item The pure hadronic model is unable to reproduce the gamma-ray and radio fluxes from the FBs
because the secondary electrons have too soft spectrum, if they
lose  energy by synchrotron radiation and inverse Compton scattering. In order to
obtain the observed radio emission in this model an additional
component of primary electrons with a hard spectrum is necessary, or very effective adiabatic losses are required.
\item The additional component of primary electrons contributes into the total gamma-ray
flux from the FBs by inverse Compton. The relation between
components produced  by protons and primary electrons depend on
the magnetic field strength in the FB and the spectral index of
primary electrons.
\item For the spectrum of primary electrons as $\propto E^{-2}$, the $pp$ collisions
collisions can only provide about  80\% of the FB gamma-ray flux in the most favorite conditions when $H\simeq 5$ $\mu$G.
\item If the magnetic field strength is larger than 7 $\mu$G then neither the
hadronic nor the leptonic models are able to reproduce gamma-ray
and radio  emission from the FBs if the spectrum of primary
electrons is $\propto E^{-2}$. For a harder spectrum, e.g. as
$E^{-1}$ a mixture of the hadronic and leptonic model is able to
reproduce the observed gamma and radio emission from the FBs even
for $H> 10$ $\mu$G.
\item With decrease of the magnetic field strength the contribution from primary electrons
into the total FB gamma-ray flux increases, and at $H\simeq 2.5$
$\mu$G the origin of gamma-rays  from the FB is pure leptonic.
\item In principle, the pure hadronic model is able to reproduce the spectra of radio and gamma-ray emission
from the FBs in the conditions of a strong plasma outflow from the
GC when the rate of adiabatic loss exceeds losses of synchrotron radiation and inverse Compton scattering.
The spectrum of secondary electrons in this case is relatively
hard, $\propto E^{-2}$ that is necessary for the observed radio
emission from the FBs.
\item However,  in this case the required values of the magnetic field strength in the FBs and the power
of CR sources are much higher than followed from observations.
 We conclude that  all versions of the hadronic model of the FBs, which we
analysed, are in our opinion problematic. In any case it is not
easy to reproduce characteristics of radio and gamma-ray emission
in the framework of this model.
\end{itemize}

\section*{Acknowledgements}

KSC is supported by the GRF Grants of the Government of the Hong
Kong SAR under HKU 701013. DOC is supported in parts by  the LPI
Educational-Scientific Complex and Dynasty Foundation. DOC and VAD
acknowledge support from the RFFI
grants 12-02-00005, 15-52-52004, 15-02-02358, 15-02-08143. CMK is
supported, in part, by the Taiwan Ministry of Science and Technology grant MOST 102-2112-M-008-019-MY3. KSC, DOC, and
VAD acknowledge support from the International Space Science
Institute to the International Team "New Approach to Active
Processes in Central Regions of Galaxies".


\begin{thebibliography}{99}
\bibitem[Ackermann et al.(2011)]{acker11}
Ackermann, M.,  Ajello, M., Allafort, A. et al. 2011, Sci, 334,
1103
\bibitem[Ackermann et al.(2012)]{acker}
Ackermann, M., Ajello, M., Atwood, W. B. et al. 2012, \apj, 750, 3
\bibitem[Ackermann et al.(2014)]{acker14}
Ackermann, M., Albert, A., Atwood W. B. et al. 2014, ApJ, 793, 64

\bibitem[Ade et al.(2013)]{ade13}
Ade, P. A. R., Aghanim, N., Arnaud, M. et al.(Planck Collaboration)  2013, A\&A, 554, 139
\bibitem[Atoyan(1992)]{atoyan}
Atoyan, A.M. 1992, A\&A, 257, 476

\bibitem[Berezinskii et al.(1990)]{ber90}
Berezinskii, V. S., Bulanov, S. V., Dogiel, V. A., Ginzburg, V.
L., \& Ptuskin, V. S. 1990, {\it Astrophysics of Cosmic Rays}, ed.
V.L.Ginzburg, (Norht-Holland, Amsterdam)

\bibitem[Blumenthal \& Gould(1970)]{blum}
Blumenthal, G. R., \& Gould, R. J. 1970, Rev. Mod. Phys., 42, 237

\bibitem[Breitschwerdt et al.(1991)]{breit91}
Breitschwerdt, D., McKenzie, J. F., \& V\"{o}lk, H. J. 1991, A\&A,
245, 79
\bibitem[Breitschwerdt et al.(2002)]{breit}
Breitschwerdt, D., Dogiel, V. A., \& V\"{o}lk, H. J. 2002, A\&A,
385, 216

\bibitem[Carretti et al.(2013)]{carr13}
Carretti, E., Crocker, R. M., Staveley-Smith, L. et al. 2013, Nature,
493, 66
\bibitem[Cheng et al.(2006)]{cheng1}
Cheng, K.-S., Chernyshov, D. O. \& Dogiel, V. A. 2006, \apj, 645,
1138
\bibitem[Cheng et al.(2007)]{cheng2}
Cheng, K.-S., Chernyshov, D. O. \& Dogiel, V. A. 2007, A\&A, 473,
351
\bibitem[Cheng et al.(2011)]{cheng} Cheng, K.-S.,
Chernyshov, D. O., Dogiel, V. A., Ko, C.-M., \& Ip, W.-H. 2011,
ApJ, 731, L17
\bibitem[Cheng et al.(2012)]{cheng12}
Cheng, K.-S., Chernyshov, D. O., Dogiel, V. A., et al. 2012, \apj,
746, 116
\bibitem[Cheng et al.(2014)]{cheng14a}
Cheng, K.-S., Chernyshov, D. O., Dogiel, V. A., \& Ko, C.-M. 2014,
\apj, 790, 23

\bibitem[Crocker \& Aharonian(2011)]{crock11}
Crocker, R.M. \& Aharonian, F. 2011, PRL, 106, 101102
\bibitem[Crocker et al.(2011)]{crocker11}
Crocker, R. M., Jones, D. I., Aharonian, F. et al.  2011, MNRAS, 413, 763
\bibitem[Crocker et al.(2013)]{crock13}
Crocker, R. M., Bicknell, G. V., Carretti, E., Hill, A., S., \& Sutherland, R. S. 2013, \apj, 791, L20
\bibitem[Dobler \& Finkbeiner(2008)]{dob08}
Dobler, G., \& Finkbeiner, D. P. 2008, ApJ, 680, 1222
\bibitem[Dobler et al.(2010)]{dob10}
Dobler, G., Finkbeiner, D. P., Cholis, I., et al. 2010, ApJ, 717,
825

\bibitem[Finkbeiner(2004)]{fink}
Finkbeiner, D. P. 2004, ApJ, 614, 186
\bibitem[Fujita et al.(2013)]{fuji}
Fujita, Y., Ohira, Y., \& Yamazaki, R. 2013, ApJ, 775, L20
\bibitem[Ginzburg(1979)]{ginz79}
Ginzburg, V. L.  {\it Theoretical physics and astrophysics}, Oxford, Pergamon Press, (International Series in Natural Philosophy. Volume 99), 1979

\bibitem[Hooper \& Slatyer(2013)]{hoop}
Hooper, D., \& Slatyer, T.R. 2013, Physics of the Dark Universe, 2,
118
\bibitem[Jones et al.(2012)]{jones12}
Jones, D. I., Crocker, R. M., Reich, W., Ott, J., \& Aharonian, F. A. 2012, ApJ, 747, L12

\bibitem[Kamae et al.(2006)]{kamae}
Kamae, T., Karlsson, N., Mizuno, T. et al. 2006, ApJ, 647, 692

\bibitem[Mertsch \& Sarkar(2011)]{mertsch}
Mertsch P. \& Sarkar, S. 2011, PhRvL, 107, 1101

\bibitem[Shibata et al.(2013)]{shiba13}
Shibata, T., Ohira, Y., Kohri, K. \& Yamazaki, R. 2014, APh, 55, 8
\bibitem[Snowden et al.(1997)]{snowden}
Snowden, S. L., Egger, R., Freyberg, M. J. et al. 1997, ApJ, 485, 125
\bibitem[Su et al.(2010)]{meng}
 Su, M., Slatyer, T. R., \& Finkbeiner, D. P. 2010, ApJ, 724, 1044
\bibitem[Thoudam(2013)]{thoudam13}
Thoudam, S. 2013, ApJ, 778, L20
\bibitem[Yang et al.(2012)]{yang12}
Yang, H.-Y. K., Ruszkowski, M., Ricker, P. M., Zweibel, E. \& Lee,
D. 2012, ApJ, 761, 185
\bibitem[Yang et al.,(2014)]{Yang2014}
Yang R.-Z., Aharonian, F., \& Crocker, R. 2014, A\&A, 567, A19

\end{thebibliography}
\end{document}